# Harmonic mesh modelling of hexagonal FeN stripes on Cu(001)

Wolfgang Kappus

Alumnus of ITP, Philosophenweg 19, D-69120 Heidelberg

wolfgang.kappus@t-online.de

## Abstract

Hexagonal FeN monolayers on Cu(001) show a stripe pattern composed of dark and bright stripes as observed recently by scanning tunneling microscopy (STM). Here a harmonic mesh model is proposed to interpret those results. It uses a planar Fourier analysis with two different wave vectors to calculate small displacements of Fe atoms caused by lattice mismatch stress. The area of displaced Fe atom triangles is used as a measure for the protrusion of N atoms in their center. Proper selection of the wave vectors allows to visualize the dominant protrusions of $(\sqrt{3} \times \sqrt{3})$ ordered nitrogen atoms. The wave vectors selected are related to Fe configurations which minimize lattice mismatch stress. The model is kept simple, so just the most basic features of the STM experiments are covered. The limits of the model are discussed and ways for extending it are proposed.

## Keywords

Cu(001) surfaces
Fe-N monolayer
Nanostripes
Lattice mismatch
Harmonic approximation
Spatially oscillating displacement fields

## 1. Introduction

This communication focuses on a recent extensive structural model of FeN stripes on Cu(001) reported by Yamada et al. [1] based on images obtained with scanning tunneling microscopy (STM). A further model is proposed explaining stripe formation adressing oscillating static strain fields generated by lattice mismatch between the hexagonal FeN monolayer and the Cu(001) substrate. For the bibliographic context reference is made to [1].

The stripe pattern observed [1] consists of dark (1x1) and bright $(\sqrt{3} \times \sqrt{3})$ patterns denoted A and B stripes. The A stripes consist of a mesh of equilateral triangles, the B stripes show an uniaxially compressed lattice of isosceles Fe triangles. In both cases N atoms sit above the triangles' centers; in the B stripe case certain N atoms are protruded, the reason for their brightness in STM experiments. Zigzag Fe rows are located within Cu(110) troughs, the stripe direction is appr. 30° off in (310) direction. B stripes show quasi-periodic faults of N atom positions. The stripe wavelength is about 35 Å [1] or 10 Cu lattice constants.

The current model is built on a mesh of Fe triangles. It uses the harmonic approximation for the effective potentials causing the small displacements of Fe atoms. Their displacement field is calculated using a plane wave expansion. Responsible for the static displacements is a mismatch between the Cu(001) substrate and the hexagonal FeN overlayer. Compression of Fe triangles creates protrusion of N atoms and is responsible for strain absorption.

After this introduction the harmonic mesh model will be detailed in section 2; starting with simple stripes the more complicated stripes of FeN on Cu(001) will follow. In section 3 the limits of such a model will be discussed and open questions will be formulated. A summary will follow as section 4.

# 2. Harmonic mesh model

A two dimensional model describes the mesh of Fe triangles with N atoms atop visible in the STM experiments [1], where A stripes consisting of equilateral Fe atom triangles are alternating with B stripes of compressed isosceles triangles. The wavelength of stripes measured orthogonal is $w_{AB} \approx 35$ Å or ca. 10 Cu(100) lattice constants. Equilateral Fe triangles within an A stripe have two atoms on one Cu(110) trough and the third on a neighboring trough, their height is $h_{Fe}$=2.65 Å. The uncompressed equilateral Fe-Fe distance $a_{Fe}$ is assumed $2*h_{Fe}/\sqrt{3}$ =3.06 Å.

Compressed isosceles Fe triangles within a B stripe again have two atoms on one Cu(110) trough and the third on a neighboring trough, their height is $h_B$=2.49 Å. The lattice constant of Cu(110) $h_{Cu}$= 2.56 Å, so the top Fe atoms in a Fe triangle differ from their "natural" location by $\delta h \approx \pm 0.08$ Å. In B stripes a subset of N atoms protrudes stronger; this is the reason for their brightness in STM experiments; this subset follows a ($\sqrt{3} \times \sqrt{3}$ ) structure with quasi-periodic faults.

## 2.1. Simple stripes

In a simple stripe model N atoms are protruded when Fe triangles are compressed. It is assumed that the location of protruded N atoms is the gravity center of compressed triangles. The regular structure of the stripes in [1] suggests a Fourier analysis, and the small deviations of the Fe-Fe distances suggest a harmonic approximation of the forces acting and the related displacements.

The mesh of Fe atoms is defined by the sum of their position $\boldsymbol{r}$ and their displacement $\boldsymbol{s}$. Fe positions within a triangular grid are

$$\boldsymbol{r}_{kl} = k*\boldsymbol{r}_1 + l*\boldsymbol{r}_2 \qquad (2.1)$$

with $\boldsymbol{r}_1=a_{Fe}(1,0)$ and $\boldsymbol{r}_2=a_{Fe}(1/2,\sqrt{3}/2)$, where $a_{Fe}$ is the unstressed Fe-Fe distance. Fe displacements $\boldsymbol{s}$ are

$$\boldsymbol{s}_{kl} = s_0(\boldsymbol{q})*\cos(\boldsymbol{q}*\boldsymbol{r}_{kl}) , \qquad (2.2)$$

where $\boldsymbol{q}$ reflects the stripe oscillations and $s_0(\boldsymbol{q})=f_0\hat{\boldsymbol{q}}$ is the unit vector $\hat{\boldsymbol{q}}$ of $\boldsymbol{q}$ multiplied with a displacement factor $f_0$. Defining the gravity center of a displaced triangle $kl$

$$\boldsymbol{d}_{kl} = \boldsymbol{r}_{kl} + (\boldsymbol{s}_{kl} + \boldsymbol{s}_{k+1,l} + \boldsymbol{s}_{k,l+1})/3 , \qquad (2.3)$$

and the triangle's area with $b_{kl}$, a graph $\{\boldsymbol{d}_{kl}\}$ can be constructed with thin points, if $b_{kl} \geq a_{Fe}^2 \sqrt{3}$ /4 (expanded triangles) and bold points, if $b_{kl} < a_{Fe}^2 \sqrt{3}$ /4 (compressed triangles).

## 2.2. Oscillation wavelengths

The wave vector $\boldsymbol{q}$ relates to the oscillation wavelengths $w_x$ in Cu(110) direction and $w_y$ in Cu(-110) direction respectively:

$$\boldsymbol{q} = 2\pi (1./w_x , 1./w_y) . \qquad (2.4)$$

From Figs. 3.a and 5 of [1] an estimate can be taken: $w_x \approx 26 \ h_{Cu}$ and $w_y \approx 15 \ h_{Cu}$. The associated graph is not presented since not relevant for the FeN-Cu(001) case.

## 2.3. Stripes with N($\sqrt{3} \times \sqrt{3}$) domains

B stripes show bright protruded N atoms in a ($\sqrt{3}$ x$\sqrt{3}$) structure where Fe triangles are strongly compressed [1]. The location of strongly protruded N atoms will be assumed on the gravity center of strongly compressed triangles. The ($\sqrt{3}$ x$\sqrt{3}$) structure implies a second oscillation. The wave vector

$$\boldsymbol{q}^{\cdot} = 2\pi\left(2/3 \, , \, -2/\sqrt{3}\right)\big/a_{\text{Fe}} \qquad (2.5)$$

of this second oscillation reflects a grid spanned by $(3/2, \sqrt{3}/2)a_{\text{Fe}}$ and $(3/2, -\sqrt{3}/2)a_{\text{Fe}}$.

Eq. (2.2) is then replaced by

$$s_{\text{kl}} = s_0(\boldsymbol{q}) * \cos(\boldsymbol{q} * \boldsymbol{r}_{\text{kl}}) + s_0'(\boldsymbol{q}^{\cdot}) * \cos(\boldsymbol{q}^{\cdot} * \boldsymbol{r}_{\text{kl}}). \qquad (2.6)$$

The construction of the graph $\{\boldsymbol{d}_{\text{kl}}\}$ needs now to be adapted to strongly compressed triangle areas and depends on the choice of the magnitude of $s_0(\boldsymbol{q})$ and $s_0'(\boldsymbol{q}^{\cdot})$. For demonstration purposes $f_0$ and $f_0'$ both are assumed $a_{\text{Fe}}/10$, nearly an order of magnitude too large. The choice of $f_0$ and $f_0'$ influences the triangle areas and thus also the widths of the A and B stripes.

Before discussing the $\{\boldsymbol{d}_{\text{kl}}\}$ graph in detail it is instructive to comment in Fig. 1 the graph $\{\boldsymbol{r}_{\text{kl}}+\boldsymbol{s}_{\text{kl}}\}$, the location of displaced Fe atoms according to Eqs. (2.1) and (2.6), bold dots, embedded in the rigid grid of Cu(001) atoms, small dots.

To distinguish the different symmetries, the hexagonal Fe atom grid is plotted as parallelogram while the Cu grid is a square. The zig-zag alignment of Fe atoms in Cu(110) troughs (caused by wave vector $\boldsymbol{q}^{\cdot}$) is evident, also the long wavelength arc (caused by $\boldsymbol{q}$). The Fe triangles are deformed, they show voids and swells. The triangles with two Fe atoms in a lower troughs and one Fe atom in an upper trough have a Nitrogen atom in their center (not shown). The smaller a triangle's area is, the higher is the Nitrogen atom protruded.

Out[70]= 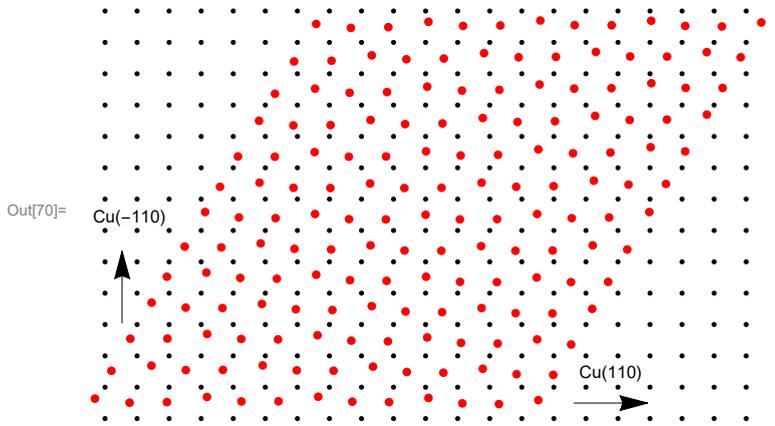

Fig. 1: Iron atoms position graph $\{\boldsymbol{r}_{\text{kl}}+\boldsymbol{s}_{\text{kl}}\}$ on Cu(001). Displaced positions of Fe atoms according to Eqs. (2.1) and (2.6): bold dots located in Cu(110) troughs. Rigid Cu grid: small dots.

The nitrogen graph $\{\boldsymbol{d}_{\text{kl}}\}$ is shown in Fig. 2. Normal N locations represented by small dots, strongly protruded N locations in B stripes represented by bold dots. The substrate directions Cu(110) and Cu(-110) are marked. The stripes show an angle of ca. -30° to the (110) direction. Strongly protruded N atoms are lined up in (-110) direction. Comparison with Figs. 1 and 3 of [1] shows agreement with exception of quasi periodic faults in (110) direction.

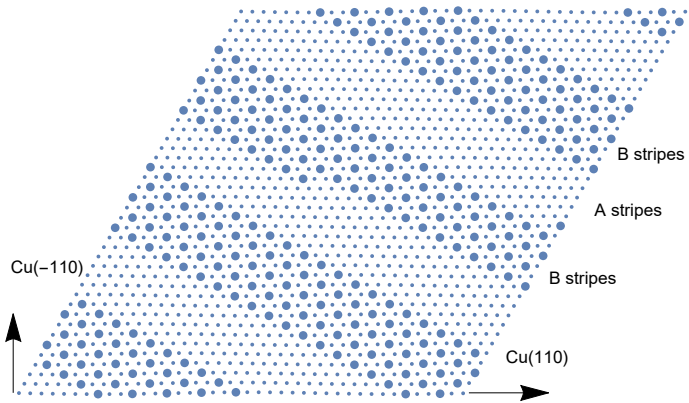

Fig. 2: Nitrogen position graph $\{d_{kl}\}$ of FeN on Cu(001). A and B stripes are oriented in (310) direction. Small dots show all N positions in the center of Fe triangles, bold dots show positions of strongly protruded nitrogen atoms in the center of strongly compressed Fe triangles within B stripes. Displaced positions of Fe atoms are according to Eqs. (2.1) and (2.6).

# 3. Discussion

The current work is intended as a step forward in the tradition of [2, 3, 4].

The current harmonic mesh model is a first approximation to the wealth of information found in [1]; it cannot be expected to cover the entire subject. An obvious lack is the micro domain issue, omitting the faults of stripe B protrusions. Also the antiphase protrusion alignment between A and A' stripes (across a B stripe) is not covered. The Fe-Fe bond probably has a strong third order component and the plane wave ansatz lacks higher frequencies.

The harmonic mesh model, however, is able to interpret key experimental data of [1] reasonably and provides a solid base for more detailed investigations.

The choice of the wave vectors $q$ and $q'$ needs explanation. The mismatch of the Cu substrate and the Fe monolayer is corrected by displacements of Fe atoms in (110) and (-110) direction. Like in the case of a rigid rod, displacements in (-110) direction allow to release stress within troughs without changing the Fe-Fe bond length significantly. This is the reason for significant (-110) components of $q$ and $q'$. The choice of $w_x$ in section 2.2 was made to fit two observations in [1]: the appr. 30° stripe angle and the appr. 15 $h_{Cu}$ wavelength in (-110) direction.

The zig-zag configuration of Fe within troughs caused the short wavelength associated with $q'$ helps to avoid Fe-Fe bond length stretching.

As next step an extension for the two possible N atom locations within a hexagonal unit cell seems desirable, utilizing a broader base of the STM data. An updated model hopefully would fit to much more details provided in [1]. The quasi periodic faults in B stripes could be related to a third oscillation. The sensitivity of stripe parameters to small changes of Cu lattice constants and to Fe-Fe bond lengths could help to determine more exact values of those parameters.

# 4. Summary

Based on an extensive structural model of FeN on Cu(001) [1] a harmonic mesh model is built for the displacements of Fe atoms. A superposition of two oscillations describes the areas of Fe triangles which is a measure for compression or expansion. The protrusion of N atoms, located in the centers of triangles, is directed by the area of triangles. The formation of stripes of less protruded N atoms and strongly protruded N atoms in a $(\sqrt{3} \times \sqrt{3})$ structure is shown and discussed. The mechanism is addressed how Fe atoms are displaced to handle the lattice mismatch to the Cu(001) surface. Shortcomings of the model are sketched and ways out are proposed.

# Author contribution

The author of this paper performed all tasks himself.

# Declaration of Competing Interest

The author declares that he has no known competing financial interests or personal relationships that could have appeared to influence the work reported in this paper.

# Acknowledgement

This work is dedicated to the F. Komori team. Many thanks to them for opening the door to this fascinating topic in the nanoworld. Many thanks also to M. Yamada for critical reading of the manuscript and for his fruitful comments and to T.&U. zur Nieden for raising helpful questions and giving kind language and style advice.